\def\be{\begin{equation}}
\def\ee{\end{equation}}
\def\ber{\begin{eqnarray}}
\def\eer{\end{eqnarray}}
\def\bers{\begin{eqnarray*}}
\def\eers{\end{eqnarray*}}

\documentclass[aps,prb,twocolumn,groupedaddress,showpacs,amsmath,amssymb]{revtex4-1}
\usepackage{dcolumn}   
\usepackage{amsmath}
\usepackage{graphicx}    
\usepackage{subfigure}
\usepackage{color}
\newcommand{\comment}[1]{}
\usepackage{ifthen}
\newboolean{includefigs}
\setboolean{includefigs}{true}      
\newboolean{includetext}
\setboolean{includetext}{true}     
\newcommand{\condcomment}[2]{\ifthenelse{#1}{#2}{}}
\begin{document}

\title{Electronic structure, magnetism and antisite disorder in CoFeCrGe and CoMnCrAl quaternary Heusler alloys}
\author{Enamullah,$^{1\dagger}$ Y. Venkateswara,$^{1\dagger}$ Sachin Gupta,$^{1,2,\dagger}$, Manoj Raama Varma,$^{3}$ Prashant Singh,$^{4}$ K G Suresh,$^{1}$ and Aftab Alam$^{1}$}
\email{aftab@phy.iitb.ac.in, enamullah@phy.iitb.ac.in;\\ $^{\dagger}$ Authors have equal contribution}
\affiliation{$^{1}$Department of Physics, Indian Institute of Technology Bombay, Mumbai 400 076, India}
\affiliation{$^{2}$Advanced Institute for Materials Research, Tohoku University, Sendai-980-8577, Japan}
\affiliation{$^{3}$National Institute for Interdisciplinary Sciences and Technology (CSIR), Thiruvananthapuram, India}
\affiliation{$^{4}$Ames Laboratory, U.S. Department of Energy, Iowa State University, Ames, Iowa 50011-3020, USA}

\begin{abstract}
We present a combined theoretical and experimental study of two quaternary Heusler alloys CoFeCrGe (CFCG) and CoMnCrAl (CMCA),
promising candidates for spintronics applications. Magnetization measurement shows
the saturation magnetization and transition temperature to be $3\; \mu_B$, $866$ K
 and $0.9 \; \mu_B$, $358$ K for CFCG and CMCA respectively. The magnetization values agree fairly well with our theoretical results and also obey the
 Slater-Pauling rule, a prerequisite for half metallicity.
 A striking difference between the two systems is their structure; CFCG crystallizes in fully ordered Y-type structure while CMCA has L2$_1$ disordered structure.
 The antisite disorder adds a somewhat unique 
 property to the second compound, which arises due to the probabilistic mutual exchange of Al positions with Cr/Mn and such an effect is possibly expected due to
 comparable electronegativities of Al and Cr/Mn. {\it Ab-initio}
 simulation predicted a unique transition from half metallic ferromagnet to metallic antiferromagnet beyond a critical excess  concentration of Al in the alloy.

\end{abstract}

\date{\today}
\maketitle
\section{INTRODUCTION}
{\par} 
Spintronics technology based on the spin degree of freedom of electrons has potential advantages over conventional electronics,
such as high speed data processing, low power consumption, 
large circuit integration density etc. and is rapidly growing.\cite{Felser}
There are many materials such as simple transition metal oxides 
(CrO$_2$, Fe$_3$O$_4$), perovskite manganites, transition metal chalcogenides, diluted magnetic semiconductors and 
many Heusler alloys(HA), which are promising for spintronics applications.\cite{Felser,Galanakis}
The striking feature of these materials is their half metallic (HM) property. From the band concept, half metallicity arises due 
to the existence of finite density of states for one spin subband (majority channel) and a
 finite band gap for the other (minority channel) at the Fermi level (E$_{\text{F}}$). The imbalance in the two
 densities of states results in 100$\%$ spin polarization of conduction (majority) electrons at E$_{\text{F}}$.
 A ferromagnetic material having
HM property is called HM-ferromagnet. Having such type of band structure in the material makes it promising for spin
injection and spin manipulation in spintronic devices. Among the systems mentioned, HA emerge out to be the most favored as HM- ferromagnets 
because of their high Curie temperature (T$_{\text{C}}$)
and structural compatibility{\cite{Far,Has1,Has2,Has3}} compared to those of conventional semiconductors.
Conventional or full HA crystallize in the ordered L2$_1$ structure with composition X$_{2}$YZ in which X and Y are the transition
metals whereas Z is a nonmagnetic element.  A new structure arises when one of X is replaced by a different transition metal, i.e., 
the stoichiometry becomes 1:1:1:1 and such alloys are 
known as quaternary Heusler alloys {\cite{Lakhan1,Lakhan2,Ozd,Ali,Dai,Sin,Ali1,Gal,Luo,Gal1,Mei}} (QHA)
with the formula XX$'$YZ. The resulting compound crystallizes in LiMgPdSn prototype structure (or Y-structure).
If Y and Z atoms randomly occupy either of the sites, the resulting structure is XX$'$Y$_{2}$/XX$'$Z$_{2}$. Such a structure is refereed to as
L2$_{1}$ disordered structure. 

Along with the theoretical prediction of halfmetallicity in HA,\cite{Gao1} a lot of
experimental work on Co-based quaternary HA has been reported.{\cite{Ali,Klae1,Alij1,LB2}}
In this regard, structural analysis, electronic and magnetic properties along with the prediction of high spin polarization in QHA have also been studied experimentally.{\cite{Alij1}} Element-specific magnetic moments and spin resolved density of states in QHA are measured using x-ray absorption spectroscopy.{\cite{Klae1}}
Spin polarization measurements in CoFeCrAl using point contact Andreev reflection (PCAR) technique
reveals 63$\%$ of spin polarized electrons at E$_{\text{F}}$.\cite{LB2} It has frequently been observed that among all the Heusler alloys, Co-based HA are the perfect materials 
for spintronic applications because of the high value of T$_{\text{C}}$ and spin polarization.

In this paper, we report a detailed theoretical and experimental study of two alloys; CoFeCrGe (CFCG) and CoMnCrAl (CMCA).
CFCG is found to have the LiMgPdSn prototype (Y-structure) with space group F$\bar{4}$3m whereas 
CMCA has L2$_{1}$ disordered structure. 
Magnetization measurement shows the saturation magnetization of 3$\mu_{\text{B}}$ and 0.9$\mu_{\text{B}}$
for CFCG and CMCA respectively, which obeys Slater-Pauling rule.{\cite{JCSlat,LPaul}} 
First principle calculation also yields the same results.
In addition, we have also studied the possible effect of antisite disorder
(L2$_{1}$ disorder) between (Mn$_{1-\text{x}}$Al$_{1+\text{x}}$) and (Cr$_{1-\text{x}}$Al$_{1+\text{x}}$) pairs in CMCA alloy.
Interestingly, a unique transition from half metallic to metallic state occurs if we go beyond 3.70$\%$ Al-excess in both
Mn-Al and Cr-Al pairs.


\section{EXPERIMENTAL TECHNIQUES AND COMPUTATIONAL DETAILS}

\subsection{Experimental Techniques}
Both the polycrystalline alloys i.e. CFCG and CMCA were synthesized by arc melting the stoichiometric amounts of constituent elements with purity of at 
least 99.99\% in water cooled copper hearth under high purity argon atmosphere. To compensate the weight loss in CMCA due to Mn evaporation, 2\% extra Mn was taken.
The formed ingots were melted several times for better mixing. As-cast samples were sealed in evacuated quartz tubes and annealed for 7 days at 800$^{0}$C 
followed by ice/water mixture quenching.
To check phase purity of samples, X-ray diffraction (XRD) patterns were taken at room temperature, using X'Pert Pro diffractometer with Cu K-$\alpha$ radiation.

XRD analysis is done with the help of FullProf suite that uses the least square refinement between experimental and calculated intensities. It contains a number of programs such as DICVOl06 for indexing XRD pattern, GFourier for calculating and visualizing electron density within the unit cell etc. for different purposes in XRD and neutron diffraction (ND) data analysis. Profile Matching well known as Lebail fitting is done by refining lattice constant, peak profile shape parameters of pseudo-Voigt function as described in FullProf manual.\cite{fullprof} GFourier program is used for the calculation and visualization of electron density within the unit cell. The visualization is very useful in identifying the atomic positions of constituent elements within the unit cell for known or unknown crystals i.e. denser the electron density contours indicate the position of heavier element among the constituent elements in the unit cell. The function to be minimized in the Rietveld Method is:

\begin{equation}
\chi^2 =\sum_{i=1}^n \ w_i \{ y_i - y_{ci} \} ^2
\end{equation}
with $w_i=1/\sigma_i^2$, where $\sigma_i^2$ is the variance of the $``$observation" $y_i$. Here $y_i$ and $y_{ci}$ are the observed and the calculated scattering intensities for a diffraction angle $2\theta_i$.\cite{fullprof} The smaller the value of $\chi^2$, the better is the refinement.

The patterns for CMCA were the same before and after annealing, but for CFCG alloy, a small amount of secondary phase was seen after annealing. 
Therefore, as-prepared CFCG and annealed CMCA were used for magnetization M(H,T) measurements.
M(H,T) was measured using a Physical Property Measurement System (PPMS) (Quantum Design). High temperature magnetization measurements were performed
with an oven attached to the PPMS.
 
As discussed in the introduction, full Heusler alloy (FHA) structure comprises of four inter-penetrating face centered cubic (fcc) sublattices and can be 
thought of as the superposition of
rock salt(NaCl) and zinc blend(ZnS) type structures.{\cite{Tanja}} In NaCl structure, each Na(Cl) atom is surrounded by six Cl(Na) atoms whereas in ZnS 
structure, each Zn is surrounded by four 
S atoms and vice-versa. The atomic sites of NaCl structure is called octahedral sites whereas the sites in ZnS are known  as tetrahedral sites. 
The ionic nature of bonds in NaCl arises due to
the large difference in the electronegativity values between the constituent elements. The covalent bonding nature arises when the 
difference in electronegativity values of the constituent elements
is very small e.g. in ZnS structure. In HA if one considers most electronegative element (usually from p-block) at (0,0,0)fcc site, (1/2,1/2,1/2)fcc site 
will be occupied by the least electronegative
element (usually low valance transition metals).{\cite{Tanja}} The remaining two fcc sublattices i.e. (1/4,1/4,1/4)fcc and (3/4,3/4,3/4)fcc sites will be occupied by the
intermediate electronegative elements among the
constituent elements. The same nomenclature for atomic sites is used here also, even though they are not surrounded by number of atoms that gave the name. 
For example, octahedral sites (0,0,0)fcc and
(1/2,1/2,1/2)fcc are surrounded by eight atoms instead of the six atoms suggested by its name. For a FHA of the type X$_2$YZ, X atoms are of the intermediate 
electronegativity values and occupy 8c(1/4,1/4,1/4),
(two fcc sublattices with atoms at (1/4,1/4,1/4) and (3/4,3/4,3/4)), Y  occupies 4a(0,0,0) (one fcc sublattice at (0,0,0)) and Z atom occupies 4b (1/2, 1/2, 1/2) 
(one fcc sublattice at (1/2,1/2,1/2)) Wyckoff 
positions of the space group Fm-3m.\cite{Tanja} The unit cell can be shifted translationally or rotationally by any amount in the crystal and its structure remains the same.
If the unit cell of the above atomic
positions is shifted by (1/2,1/2,1/2), new atomic positions will be X at 8c, Y at 4b and Z at 4a Wyckoff sites. There can be other similar combinations. 
For the case of QHA, if Z atom is considered at 4a (0,0,0)
position, the remaining three atoms X, X$'$ and Y will be placed in three different fcc sublattices 4b(1/2,1/2,1/2), 4c(1/4,1/4,1/4) and 4d(3/4,3/4,3/4) in three
non-degenerate ways such that there are only three
independent atomic arrangements in the Y structure. As discussed, any translation of the unit cell does not change the crystal structure, and shifting of these 
configurations by (1/4,1/4,1/4), (1/2,1/2,1/2) or 
(3/4,3/4,3/4) of unit cell will simply change the  origin of the atoms but not the configuration. Three configurations are shown for CFCG in Fig.1. For example, 
if the atomic positions of Cr and Co are interchanged
in Fig.1(a), the resulting structure is same as the initial one because if that primitive cell is inverted along the body diagonal, the atomic arrangements will 
be same as the initial one. These are energetically
non-degenerate configurations. Similarly the other two configurations in Fig.1 can be understood. In this way there are only three non-degenerate (distinct) atomic 
arrangements for XX$'$YZ type quaternary Heusler alloy.
The structure factor for quaternary Heusler alloy XX$'$YZ, having Z at 4a(0,0,0), Y at 4b(1/2,1/2,1/2), X at 4c(1/4,1/4,1/4) and X$'$ at 4d(3/4,3/4,3/4) is given below,
\begin{equation}
F_{hkl} = 4(f_{z}+f_{y} e^{\pi i (h+k+l)}+f_{x} e^{\frac{\pi}{2} i (h+k+l)}
+f_{x'} e^{\frac{-\pi}{2} i (h+k+l)})
\end{equation}
with unmixed (hkl) values. Here $\text{f}_{\text{x}}$, $\text{f}_{\text{x$'$}}$, $\text{f}_{\text{y}}$, and $\text{f}_{\text{z}}$ 
 are the atomic scattering factors for the atoms X, X$^{'}$, Y and Z respectively. Therefore,
\begin{equation}
F_{111} = 4\left[(f_{z}-f_{y})-i(f_{x}-f_{x'})\right]
\end{equation}
\begin{equation}
F_{200} = 4\left[(f_{z}+f_{y})-(f_{x}+f_{x'})\right]
\end{equation}
\begin{equation}
F_{220} = 4\left[(f_{z}+f_{y})+(f_{x}+f_{x'})\right],
\end{equation}
are used to classify the ordering of the crystal structure.

\begin{figure}[t]
\begin{centering}
\includegraphics[scale=0.95]{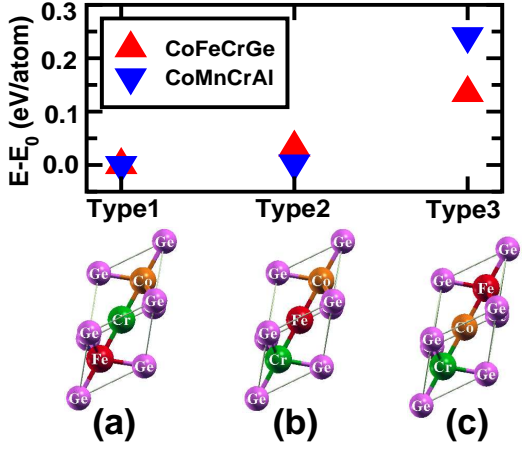}
\caption{Site preference energy plot for the different configurations of CFCG(triangle up) and CMCA(triangle down).
E$_{0}$, reference energy corresponds to Type1 structure.
Primitive unit cells (a), (b) and (c) are three non-degenerate configurations of CFCG corresponding to Type1, Type2 and Type3.}
\label{fig:difcrcfcg}
\end{centering}
\end{figure}
\subsection{Computational Details}
First principle calculations were done using a spin polarized density functional theory (DFT) implemented within Vienna ab-initio simulation package(VASP){\cite{Kres1}}
with a projected augmented-wave basis.{\cite{Kres2}}
We used Perdew-Bueke-Ernzerhof (PBE) for the electronic exchange-correlation functional.  24$^{3}$ k-mesh were used for Brillouin zone integration. A plane wave cut-off 
of 288 eV with the energy convergence criteria of 0.1 meV/cell.
In order to study the effect of antisite disorder in CMCA, we use a 3$\times$3$\times$3 super cell involving 108 atoms/cell with 27 atoms of each kind. Guided by the 
experimental findings, two types of antisite disorder were investigated i.e.
between Mn and Al (CoMn$_{1-x}$CrAl$_{1+x}$) and Cr and Al
(CoMnCr$_{1-x}$Al$_{1+x}$). Stability of such antisite disorder was checked by calculating the formation energy
 ($\Delta$E) as defined below for a general ABCD alloy,
\begin{eqnarray}
\Delta E &=& E[\text{A}_{1-x}\text{B}_{1+x} \text{C}\text{D}] -
 \left[\rule{0mm}{3.5mm}2(1-x)\ E(\text{A}_2\text{C}\text{D}) \right. \nonumber\\
&&\left. +\ 2(1+x)\ E(\text{B}_2\text{C}\text{D}) \rule{0mm}{3.5mm}\right].
\end{eqnarray}

\begin{figure}[t]
\begin{centering}
\includegraphics[scale=1.2]{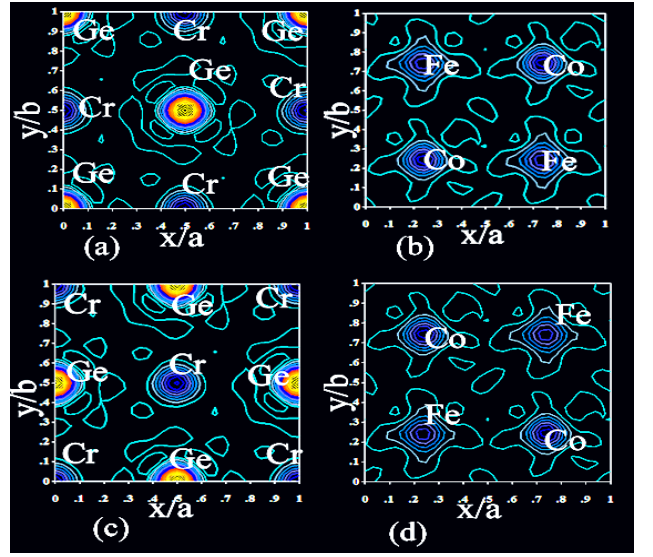}
\caption{Electronic density of individual atoms in the unit cell of CFCG at (a) z=0.0/c (b) z=0.25/c (c) z=0.5/c and (d) z=0.75/c plane.}
\label{fig:fouriermaps1}
\end{centering}
\end{figure}

\begin{figure}[b]
\begin{centering}
\includegraphics[width=0.43\textwidth, height=0.45\textwidth]{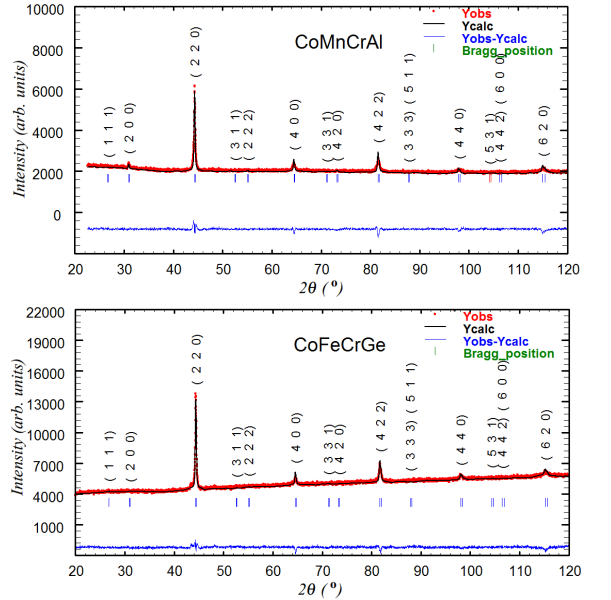}
\caption{Rietveld refinement of XRD data of CMCA(top) and CFCG(bottom).}
\label{fig:refimentCFCG}
\end{centering}
\end{figure}

\section{RESULTS AND DISCUSSION}
\subsection{Experimental}
{\par}
Rietveld refinement of powder X-ray diffraction(XRD) data using FullProf suite reveals that both CFCG and CMCA 
crystallize in the cubic structure with 
lattice constants 5.77$\pm$0.01\AA\
and 5.76$\pm$0.01\AA\ respectively. 

$\chi^2$ values of Rietveld refinement for three distinct atomic arrangements (as depicted in Fig. \ref{fig:difcrcfcg}) are presented in Table \ref{tab:chi2} for both the alloys.
\begin{table}
\caption{$\chi^2$ values of the Rietveld method for three distinct atomic arrangements (from Fig. \ref{fig:difcrcfcg}) for CFCG and CMCA}
\begin{tabular}{|r|c|c|l|}
\hline
Alloy/ Configuration & Type1 & Type2 & Type3 \\
\hline
CFCG & $\chi^2 = 2.04$ & $\chi^2 = 2.22$ & $\chi^2 = 2.29$ \\ 
\hline
CMCA & $\chi^2 = 1.72$ & $\chi^2 = 1.72$ & $\chi^2 = 1.99$\\
\hline 
\end{tabular}
\label{tab:chi2}
\end{table}

For CFCG, the constituent elements are nearest neighbors in the periodic table,
due to which their atomic scattering factors are nearly identical.
Hence the intensities of superlattice peaks (111) and (200) are very small in comparison to that of (220) peak. 
This can be understood from Eq.(1).
Therefore, one can do the refinement with all the three configurations shown in Fig.\ref{fig:difcrcfcg}
and can be fitted to XRD data. The best fit between observed and calculated intensities is observed for the first configuration. 
In this configuration, constituent elements, Ge and Cr are at 4a(0,0,0)
and 4b(1/2,1/2,1/2) octahedral sites whereas Co and Fe are at 4c(1/4,1/4,1/4) and 4d(3/4,3/4,3/4) tetrahedral sites respectively.\cite{Tanja}
 For CFCG, Cr is the least electronegative (1.66 Pauli units) \cite{wiki} and therefore, it forms ionic type sublattice 
 with Ge (which has more electronegativity of 2.01 Pauli unit) and becomes stable by donating its electrons to other 
 elements in the alloy. Ge tries to accept electrons 
 from other elements. As a result, the electronic density at Cr site decreases whereas it increases at Ge site.  
 The X and X$'$ atoms(here Fe and Co) have
intermediate electronegativities and occupy tetrahedral sites.\cite{Tanja} 
The electronic densities of various atoms in the unit cell can be visualized from the contour plot shown in Fig.\ref{fig:fouriermaps1}
generated from XRD refinement.
It is clear from Fig.\ref{fig:fouriermaps1}(a) and (c) that most of the charge is distributed around Ge atomic site,
while Cr site has the least density. Fe and Co are surrounded by intermediate charge in comparison to Cr and Ge sites in Fig.\ref{fig:fouriermaps1}(b)
and (d). This configuration is energetically most favorable as found from our calculation. 
Therefore, crystal structure shown in Fig.\ref{fig:difcrcfcg}(a) is the most stable.

\begin{figure}[t]
\centering
\includegraphics[scale=0.88]{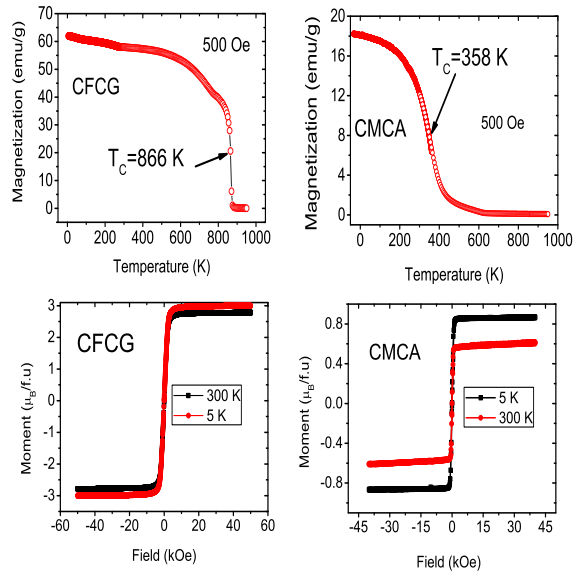}
\caption{(Top) Temperature dependence of magnetization M at 500 Oe. T$_\text{C}$ is calculated from the minima of the first order derivative of M vs. T curve.
(Bottom) Magnetic moment vs `H' at 300K and 5K
for CFCG(left) and CMCA(right).}
\label{fig:mtmag}
\end{figure}

 For CMCA, the superlattice peak (200) is more intense in comparison to (111) peak 
 and is clearly visible in the XRD pattern (Fig \ref{fig:refimentCFCG}). 
 This suggests that there is considerable amount of disorder between octahedral sites. 
 This is like B2 disorder in X$_2$YZ HA, but in QHA
 it should not be treated as B2 disorder because X and X$'$ are different atoms. 
  It is rather an L2$_1$ type disorder where (0,0,0) and (1/2,1/2,1/2) fcc sublattices are randomly occupied by the Z and Y atoms.
 From Eq.(1), it is clear that $|F_{111}|^2$ reduces as compared to $|F_{200}|^2 $ because $f_z=f_y$ due to equal probability of finding the Z and Y atoms at those sites.
 Hence the intensity, which is proportional to $|F_{111}|^2\rightarrow 0$ when $|f_x-f_{x'}|\rightarrow 0 $. Here $|f_x-f_{x'}|\approx 0$ as
 X and X$'$ are nearest neighbors. Similar
 to CFCG (Fig.\ref{fig:difcrcfcg}), CMCA also has three different configurations with the exception that there is a probability of exchange of
 atoms between the octahedral sites.
 i.e., (0,0,0) and (1/2,1/2,1/2)fcc sublattices. Even though XRD can be fitted with all the three configurations, the configuration in which octahedral site
 (1/2,1/2,1/2)fcc contains the least electronegative element is energetically favorable. Here Mn and  Cr have least electronegativity and hence the 
 two configurations (Fig1(a) 
 and (b)) containing Cr or Mn at
 (1/2,1/2,1/2)fcc sublattice are favorable. Electronegativity of Al is also of  the same order as that of  Mn or Cr and consequently Al also tries to
 occupy at (1/2,1/2,1/2)fcc. 
 As such, Al occupies different octahedral sites (0,0,0)fcc and (1/2,1/2,1/2)fcc.
  As a consequence the atoms which were initially at (1/2,1/2,1/2)fcc sites occupy both octahedral sites randomly like Al. Due to this behavior,
  (111) peak vanishes in XRD.
  Al atoms occupying  (1/2,1/2,1/2)fcc try to lose electrons. Fig.\ref{fig:refimentCFCG}(b) shows the XRD refinement by considering equal probability of finding
Mn or Al atoms in octahedral site. The conventional unit cell is shifted by (1/4,1/4,1/4) 
while doing the refinement and so the space group changes to Fm$\bar{3}$m (\# 225).  In this space group,
 the occupancies are Co at 4a, Cr at 4b and Mn/Al at 8c Wyckoff sites. 
 As Mn and Cr are neighboring elements, swapping of these elements will not be distinguishable from XRD. Due to this reason, $\chi^2$ value is same for the first and second configurations, as seen in Table \ref{tab:chi2}. 
 Therefore, the conclusion is that CFCG is fully ordered while CMCA has L2$_{1}$ disordered structure.
 
 It is observed in HA that if more than one atom has nearly same electronegative values, some degree of disorder can be expected. For example, CoMnCrAl, 
 CoFeCrAl\cite{Lakhan2} and
 Co$_{2}$Cr$_{1-\text{x}}$Fe$_{\text{x}}$Al\cite{spin-text} HA have disorder between Cr and Al sites. Disorder in these systems arises because of
 the same electronegativity values of 
 Cr and Al atoms. Consequently Al atom acts as an electron donor and occupies one of the octahedral sites
 (1/2,1/2,1/2)fcc with almost same probability of occupancy as that of Cr atoms. Similarly Mn$_{2}$CoAl \cite{YZh}
  and Co$_{2}$MnAl \cite{Vin} have a disorder between Mn and Al sites. This type of disorder is seen in HA containing Zn as well. Zn also tries
  to occupy both octahedral sites 
  (1/2,1/2,1/2)fcc and (0,0,0)fcc sublattice, because Al and Zn atoms have the nature that in some cases they lose electrons and in some other
  cases they accept electrons because of
  their low electronegativity and proximity to the p-block of the periodic table. However one can also synthesize perfectly ordered systems in
  HA containing Al atoms, such as 
  CoFeTiAl\cite{spin-text} (Y structure), Co$_2$TiAl\cite{WZhang}(L2$_1$ structure). The scenario is a bit different in this case. Since Ti atom 
  has the least electronegativity
  among the constituent atoms, behaves as a charge donor and tries to occupy the (1/2,1/2,1/2)fcc and does not allow Al to occupy the same site. 
  Hence there will be no disorder in these systems.
  Therefore, on the basis of data available on a number of alloys, we could propose an empirical relation between relative electronegativity 
  values and the occurrence of disorder. 
 
Top plots of Fig. \ref{fig:mtmag} show the temperature (T) dependence of magnetization in constant field of 500 Oe for
CFCG (left) and CMCA (right) showing the ferro-paramagnetic transition.
The Curie temperature has been determined by taking the minima of the first order derivative of Magnetization vs Temperature (M-T) curve.
The estimated T$_\text{C}$ values are about 358K and 866K for CMCA and CFCG respectively.
High T$_\text{C}$ of these alloys enable them to be potential candidates for room temperature applications.

\begin{figure}[t]
\centering
\includegraphics[width=0.48\textwidth, height=0.41\textwidth]{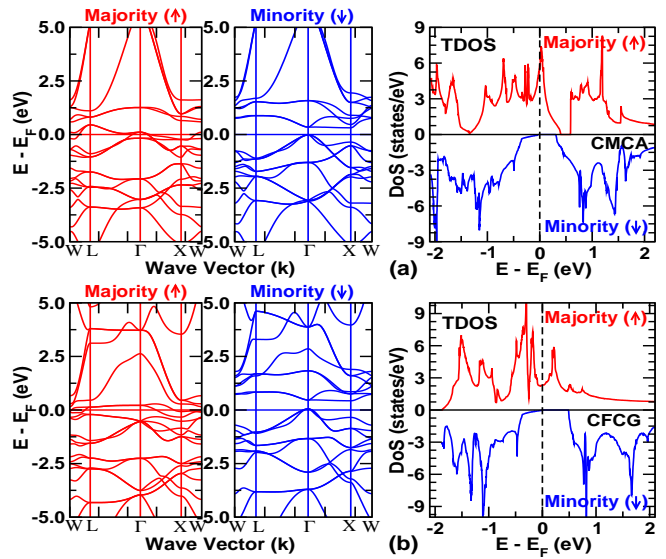}
\caption{Spin resolved band structure(left) and density of states (right) for CMCA(a) and CFCG(b) at experimental lattice constant($a_{\text{exp}}$).
Both systems clearly show half metallic behavior with a band gap $\sim$0.328 eV for CMCA and $\sim$0.481 eV for CFCG.}
\label{bs+dos_CMCA+CFCG}
\end{figure}

\begin{figure*}[t]
\centering
\includegraphics[scale=0.58]{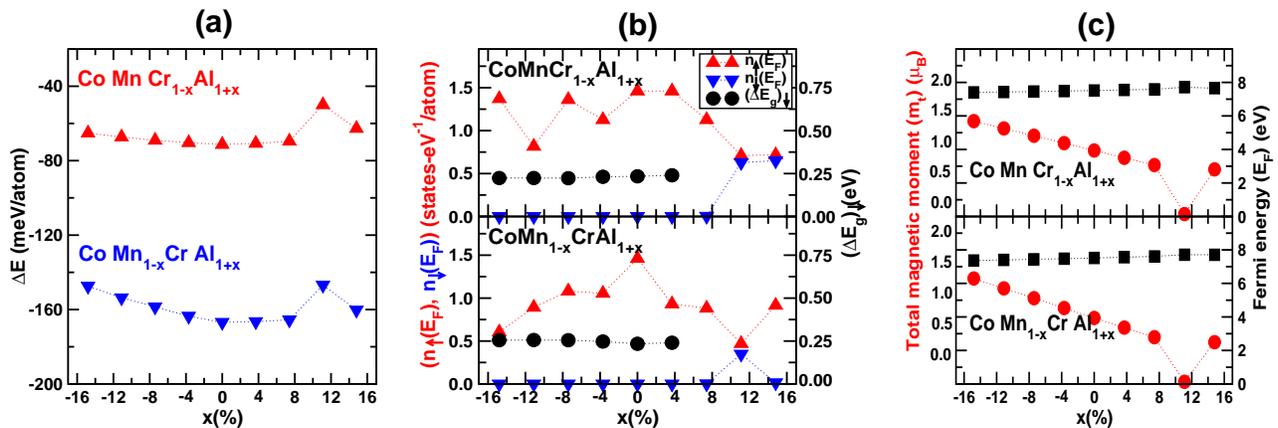}
\caption{(a) Formation energy ($\Delta$E) vs. antisite disorder (x)  for CoMnCr$_{1-\text{x}}$Al$_{1+\text{x}}$ (triangle UP)
and CoMn$_{1-\text{x}}$CrAl$_{1+\text{x}}$ (triangle DN) (b)
Concentration (x) dependence of DoS at E$_{\text{F}}$ for majority spin (triangle UP), minority spin (triangle DN) and band gap
($\Delta$E$_\text{g}$)$_{\downarrow}$ (circle) for 
CoMnCr$_{1-\text{x}}$Al$_{1+\text{x}}$ (top) and  CoMn$_{1-\text{x}}$CrAl$_{1+\text{x}}$ (bottom) (c) x-dependence of total magnetic 
moment (m$_\text{t}$) and change in Fermi energy (E$_{\text{F}}$) for the same two alloying. }
\label{disorder-fig}
\end{figure*}

Figure \ref{fig:mtmag} (bottom) shows the field dependence of magnetization for the two alloys. The absence of hysteresis reveals the soft
magnetic nature of the alloys. Both the alloys
show saturation at 5K and 300K. The saturation moment at 5 K is estimated to be 3 $\mu_B$ and 0.9 $\mu_B$ for CFCG and CMCA respectively.
The total moment in Heusler compounds can be estimated from the Slater-Pauling rule by counting the number of valence electrons in the primitive cell.{\cite{IGal}}
In QHA, the total moment (m) per unit cell can be expressed as \cite{Ali}
\begin{equation}
m=(N_v-24)\mu_B
\label{equ:muB}
\end{equation}where $N_v$ (s,d electrons for transition metals and s,p electrons for main group element) is the number of valence electrons per unit cell. 
As CFCG and CMCA have 27 and 25
valence electrons respectively, according to Slater-Pauling rule (using Eq.\ref{equ:muB}), the moment in these compounds should be 3 and 1 $\mu_B$. 
But experimentally observed magnetic
moment for CMCA (0.9$\mu_\text{B}$) slightly deviates from the Slater-Pauling rule, because of the
presence of disorder. On the other hand, in CFCG, the agreement is very good. 
In addition to the experiment, the theoretically calculated moments also agree fairly well with the Slater-Pauling prediction (described in the next section). 

\subsection{Theoretical}
{\par}
To check the stability, we have first calculated the site preference energies for various atomic configurations.
Considering the symmetry of the XX$'$YZ structure, we fix the Z-atom at 4d position and permute rest three atoms on 4a, 4b and 4c Wyckoff sites.
Out of six possible configurations, 
only three are
energetically non-degenerate, namely Type1, Type2 and Type3 as shown in the Fig.\ref{fig:difcrcfcg} for both
CFCG and CMCA. Type1 (where X atom sits at 4a, X$'$ at 4b and Y at 4c) is found to be energetically the most stable
configuration, as also configured by experiment. 

Figure \ref{bs+dos_CMCA+CFCG} shows the spin polarized band structure and density of states (DoS) for CMCA(top) and CFCG(bottom)
respectively. Half metallicity is obvious in both the systems with a finite state (at E$_{\text{F}}$) in majority channel
but gapped in minority. Calculated magnetic moment for CMCA is 0.98 $\mu_\text{B}$ ($\mu_\text{expt}$ = 0.9 $\mu_\text{B}$)
while for CFCG is 2.99 $\mu_\text{B}$ ($\mu_\text{expt}$ = 3.0 $\mu_\text{B}$), which follows the Slater-Pauling rule.

\par
Intrinsic defects such as antisite disorder is fairly common in QHA. Our XRD data clearly indicate the signature of
L2$_{1}$ disorder in CMCA, where Al site is expected to mix with Mn (and possibly with Cr). Electronic structure of any material is 
extremely sensitive to such defects, and has not received much attention in the literature. 
We have performed first principle calculation to check the stability, electronic structure and magnetism for two sets of 
antisite disorders namely CoMn$_{1-\text{x}}$CrAl$_{1+\text{x}}$ and CoMnCr$_{1-\text{x}}$Al$_{1+\text{x}}$. These are done
by using a 3$\times$3$\times$3 supercell of the primitive $4$-atom cell.

Figure \ref{disorder-fig}(a) shows the formation energy ($\Delta \text{E}$) of 
CoMn$_{1-\text{x}}$CrAl$_{1+\text{x}}$ (triangle down)
and CoMnCr$_{1-\text{x}}$Al$_{1+\text{x}}$ (triangle up) for both excess (positive x-vale) and deficit (negative x-value) of Al in the compound.
Negative values of $\Delta \text{E}$ indicates that Al indeed prefers to mix with Mn and Cr. Mn is relatively much more preferable 
to mix due to a larger negative $\Delta \text{E}$, as also revealed by our XRD data. Detailed analysis of such antisite disorder
can be accurately probed with neutron diffraction experiment.

Figure \ref{disorder-fig}(b) shows the value of DoS at E$_{\text{F}}$ for majority n$_{\uparrow}$ (triangle up) and minority 
n$_{\downarrow} $(triangle down) spin channels. The associated band gap 
$(\Delta \text{E}_{\text{g}})_{\downarrow}$ in the minority spin is also represented
(solid circle). Top (bottom) panel are the results for CoMnCr$_{1-\text{x}}$Al$_{1+\text{x}}$ (CoMn$_{1-\text{x}}$CrAl$_{1+\text{x}}$). 
Interestingly, deficit of Al (negative x) up to x $\simeq$ 14.81$\%$ maintains the half metallicity, however excess of Al (positive x) causes a 
transition from half metallic to metallic beyond x $\simeq$ 3.70$\%$ in both the cases. At 7.41$\%$ excess Al, the minority spin
tend to have a small DoS at E$_{\text{F}}$; n$_{\downarrow}$(E$_{\text{F}}$) $\simeq$ 0.03 states/eV-atom
(CoMnCr$_{1-\text{x}}$Al$_{1+\text{x}}$) and n$_{\uparrow}$(E$_{\text{F}}$) $\simeq$ 0.02 states/eV-atom
(CoMn$_{1-\text{x}}$CrAl$_{1+\text{x}}$). Such transition is something unique and has never been observed before.

It turns out that this metallic transition is intimately connected with a magnetic transition, where the system goes from
a ferromagnetic state to an antiferromagnetic state. This is shown in Fig.\ref{disorder-fig}(c), where the total magnetic moment 
changes discontinuously at the same concentration ($\text{x}\sim$7.41\%) at which the system
loses its halfmetallicity. E$_{\text{F}}$ almost remains unchanged with varying x (square symbol). Although we 
have theoretically studied the effect of antisite disorder up to $\text{x}\sim$14.81$\%$, such a large disorder may not be expected to survive in the 
actual sample. 

\section{Conclusions}
{\par}
In conclusion, CFCG and CMCA are found to be two interesting materials, the former crystallizes in Y type structure while the latter 
shows an L2$_{1}$ disordered structure, which
is due to the random occupancy of octahedral site atoms Al with Cr/Mn. Both the alloys show half metallic ferromagnetic behavior with a 
specific site preference for the constituent atoms.
CFCG is more useful because of its high Curie temperature (866 K) while CMCA shows an intrinsic antisite disorder which allows a larger
tunability of its properties. Magnetization measurement 
yields magnetic moments which obey the Slater Pauling rule, and which also agree with our theoretical prediction in both the cases.
{\it Ab-initio} electronic structure simulation confirms the 
stability and half metallicity in both the compounds. L2$_1$ disorder in CMCA is further investigated by simulating antisite 
disorder which also indicates the possibility of halfmetallic ferromagnetic
behavior in presence of small disorder. However, it changes to a metallic antiferromagnetic state beyond a certain excess Al in the alloy.

\section{Acknowledgements} 
Enamullah acknowledges IIT Bombay for providing financial assistance to carry out
postdoctoral research. PS would like to thank U.S. Department of Energy (DOE), Office of Science, Materials Science and Engineering Division for support.


\end{document}